\documentclass[12pt,american]{iopart}
\usepackage[T1]{fontenc}
\usepackage[latin1]{inputenc}
\usepackage{iopams}
\usepackage{babel}
\usepackage{setstack}
\usepackage{psfrag}
\usepackage{graphicx}

\begin{document}

\begin{flushright} {\footnotesize ESI/1611}  
\end{flushright}
\vspace{5mm}
\vspace{0.5cm}

\title{The Classical Stability Of The Ghost Condensate}

\author{Alexey Anisimov \footnote{anisimov@mppmu.mpg.de}}
\address{Max-Planck-Institut für Physik, Fohringer Ring 6,
D-80805, Munich, Germany\\
and \\
Institute of Theoretical and Experimental Physics,
117218, B.Cheremushkinskaya 25, Moscow, Russia}
\author{Alexander Vikman \footnote{vikman@theorie.physik.uni-muenchen.de}}
\address{Arnold-Sommerfeld-Center for Theoretical Physics, Department für Physik, Ludwig-Maximilians-Universität München,
Theresienstr. 37, D-80333, Munich, Germany }


\begin{abstract}
{\small We discuss classical stability of the cosmological mechanism which is responsible  
for the ghost condensation. We show that the simplest
general covariant effective action which includes two covariant derivatives may
lead to an overshoot of the condensation point into the regime, where the model 
becomes classically unstable. The model exhibits this behaviour for the de Sitter 
and matter dominated universes when the initial 
values taken to be in the region of  validity of the low-energy effective theory. 
In the matter dominated case there is a finite time (which can be large 
enough to exceed the present age of the Universe) during which the model remains
in the stable region. In inflationary regime the system does not experience
an overshoot only if $H_{inf}\gtrsim M$, where $H_{inf}$ is the Hubble
parameter during inflation, and $M$ is the characteristic scale 
of the effective field theory (EFT). This latter regime is, however, inconsistent with the 
EFT description at the condensation point. We discuss the limitations of the use of the truncated action
and point out how one can avoid problems due to the overshoot in order
to have a viable ghost condensation mechanism.}{\normalsize \par}
\end{abstract}
\maketitle

\section{Introduction}

Since the work \cite{k_inf} various theories \cite{k_ess1,k_ess2,k_ess3}
with Lagrangians $\mathcal{L}(\phi,\partial\phi)$ being arbitrary
functions of $\phi$ and $\partial_{\mu}\phi$ were explored in attempt
to address such issues as the origin of the early inflation and the
origin of the dark energy which is observed to be dominant at present.
The study of such Lagrangians is commonly motivated by the argument
that they could simply be low energy outcomes from string/M theory.
It is possible that these effective Lagrangians include also
higher order derivative terms. With a Lagrangian which depends only
on $\phi$ and $\partial_{\mu}\phi$, the usual hydrodynamic intuition
is useful, with such important quantities being well defined as the
sound speed $C_{s}$ and the equation of state $w$. Such cosmological
fluids lead to additional contributions to the energy density of the
Universe and modify the evolution of the Universe. However, the possibility
of a consistent gravity modification in the infrared using Lagrangians
with unconventional kinetic terms was generally overlooked.

Coming from another direction one can try to make various models for
direct gravity modification in order to understand the present acceleration
phase of the Universe (see, e.g. \cite{br_w,br_w1,br_w2,mod,mod1,mod2}).
Most of the models are confronted with various internal theoretical
problems such as the appearance of ghost degrees of freedom or strong
dynamics on the intermediate distances.

Recently a beautiful idea based on Lagrangians with unconventional
kinetic terms but aimed at a modification of gravity in the infrared
was proposed in \cite{ah}. It turned out to be completely free of
the usual problems which arose in other models of infrared gravity
modification. Despite a resemblance to other theories with unconventional
Lagrangians $\mathcal{L}(\phi,\partial\phi)$, the physics content is
quite different. We will briefly explain that below.

The work \cite{ah} considered an effective Lagrangian of the following
form,\begin{equation}
\mathcal{L}=M^{4}\left[P(X)+Q(X)S\left(\frac{\Box_{g}\phi}{M^{3}}\right)\right],\label{eq:Lagrangian for fi}\end{equation}
 where $M$ is the characteristic scale of the effective theory (of order of the ultraviolet (UV) cutoff), 
the dimensionless variable $X$ is defined as \begin{equation}
X\equiv\frac{1}{2}g^{\mu\nu}\left(\frac{\partial_{\mu}\phi}{M^{2}}\right)\left(\frac{\partial_{\nu}\phi}{M^{2}}\right),\end{equation}
 and $P(X)$, $Q(X)$, $S\left(\frac{\Box_{g}\phi}{M^{3}}\right)$
are some general functions, $\Box_{g}\equiv g_{\mu\nu}\nabla^{\mu}\nabla^{\nu}$
(where $\nabla^{\mu}$ is the covariant derivative). \\

For a moment we will neglect the second term in the Lagrangian. At
this point the theory we consider is a kind of k-essence theory. In
the case when $P(X)$ has a minimum at some $X^{*}\ne0$ the value
of $X$ is driven by the Hubble friction to $X=X^{*}$. We will further
refer to $X^{*}$ as the ''attractor point''.
There is a preferred reference frame where the field $\phi$ is spatially 
homogeneous and isotropic. In this reference frame the field $\phi$ is 
driven to the nontrivial background solution $\phi=M^{2}\sqrt{2X^*}t$. Thus, all this can be viewed
as a formation of Lorentz violating cosmological condensate. It is
easy to see that the reference frame where this condensate is homogeneous and isotropic
coincides with the cosmological reference frame. Up to this point
the theory is a pure k-essence model. It is easy to check that excitations
$\pi$ defined as $\phi=M^{2}\sqrt{2X^{*}}t+\pi$ do not have a $(\overrightarrow{\nabla}\pi)^{2}$-
term but only a $\dot{\pi}^{2}$- term in the Lagrangian. The $\pi$
excitations couple to gravity in a different way then a usual scalar
field. They mix with gravity and give the graviton ''mass term'',
which is described by

\begin{equation}
\delta S\propto\int d^{4}x\, M^{4}(h_{00}-\dot{\pi})^{2}.\label{eq:deltaS}\end{equation}
But since $\pi$ is not a dynamical degree of freedom at this point,
Eq. (\ref{eq:deltaS}) does not change the graviton propagator yet. 

In order to modify the graviton propagator we need to make $\pi$
dynamical. The simplest form of the second term in Eq.~(\ref{eq:Lagrangian for fi})
which yields the dynamics to $\pi$ can be chosen as $M^4 Q(X)S(\Box_{g}\phi/M^{3})=-
\alpha^{2}(\Box_{g}\phi)^{2}/2M^{2}$. We will further refer to this term as $\alpha$-term.
 If the gravity is switched off, the scalar field $\pi$ attains the
peculiar Lorenz violating dispersion relation $\omega^{2}\propto\vec{k}^{4}/M^{2}$.
The idea that a varying scalar obeys a Lorentz-breaking dispersion
relation and may induce Lorentz symmetry violation also in other sectors
of the theory (via its gradient) was also explored in \cite{other}.
If one includes gravity, $\pi$ mixes with the graviton and modifies
the graviton propagator. This may be thought of as a gravitational
analog of the Higgs mechanism. The graviton acquires a third degree
of freedom at the price of breaking the time diffeomorphism invariance.
Bizarre gravitational effects due to this modification were first
discussed in \cite{ah}. 

The energy density of the condensate along with its pressure can be
chosen by hand to be either zero or nonzero. The first case is quite
interesting because the condensate does not contribute to the Einstein
equations. But excitations of such background still modify gravity
propagation and the condensate breaks Lorenz invariance even 
in an approximate Minkowski spacetime. In the second case this contribution is not zero and
one ends up with an inflationary regime \cite{ah2}. Here we will
consider only the first case $\varepsilon(X^{*})=0$ because it is
already interesting enough and yields nontrivial physics. 

The scalar field introduced in \cite{ah} is ghost-like, i.e. the function $P(X)\propto-X$
at small $X$. But in a new nontrivial background to the right side
of $X^{*}$ the excitations $\pi$ have a regular form kinetic term.
It was argued in \cite{ah} that once the system reaches a small neighborhood
to the right of $X^{*}$ the theory is healthy in both cases: with
or without gravity. It is described by the effective Lagrangian with
all the operators behaving well in the infrared. 
As was discussed in \cite{ah} at the condensation
point terms with higher time derivatives scale away faster then those
with spatial derivatives and, thus, can be dropped out from the low
energy effective action. If one also assume that the Hubble constant
is much lower then the UV cutoff scale, one can further drop terms
like $H\dot{\phi}$ out of a $\Box\phi$ since these terms will modify
low energy effective action at the condensation point with a small
prefactor $H/M$.  

However, if one starts far from condensation point
the scaling argument is not valid anymore and these terms are important
for the evolution of the background field until the system reaches
point $X=X^{*}$. 

The details of this evolution are important because condensation point
is special. This manifest itself in the fact that the sound speed
becomes imaginary at $X<X^{*}.$ The equation of motion changes its
type from hyperbolic at $X>X^{*}$ to elliptic at $X<X^{*}.$ This,
in turn, means that one can not consider Cauchy problem in the usual
sense to the left side of the condensation point. According to Cauchy-Kowalewski
theorem it is possible to guarantee the existence and uniqueness of
the solution only in the small vicinity of Cauchy hypersurface. But
in general, finding a solution of Cauchy problem to the elliptic equation
is problematic because the solutions, as a rule, do not continuously
depend on the initial data (see, for example, \cite{vlad}). 

The above logic implies that for dynamical evolution of the background
homogeneous field it is important not to overshoot the point $X=X^{*}$. Otherwise,
one will have to cope with an unplausible elliptic regime. This instability
is different from Jeans type instability which arises due to mixing with gravity
found in \cite{ah}. The Jeans instability is physical and does not
lead to the change of the type of the equation of motion. While the
instability due to overshoot of the condensation point is, in fact,
much worse. In the elliptic regime one can not satisfy causality principle
because the Cauchi problem is commonly ill defined in this case. 

In this work we attempt to make the discussion of
the formation of the ghost condensate more detailed. We will concentrate
on a narrow scope of questions related to the dynamics which describes
its formation. In the second section we review some of the results of \cite{ah}.
In section 3 we will write the equation of motion including the
$\alpha$ term. We will argue that an overshoot of the attractor
point $X^{*}$ may occur, upsetting the classical stability of the
$\pi$ excitations. The authors of \cite{ah} considered general covariant
form for the action but showed that only asymptotically the coefficient
in front of a $(\nabla\pi)^{2}$ term approaches zero because of the
Hubble friction. In general, as we will see later, the system will
exhibit oscillatory behaviour around condensation point. These oscillations
are naturally in the UV regime. 

One can argue that from the point of view of the low energy effective
field theory alone these oscillations signal that the overall consideration
may be inconsistent and one have to include an infinite number of
higher dimensional operators. However, if one wants to have a viable
dynamical mechanism it is necessary to ensure that starting from a general
covariant action the evolution of the background field has to be smoothly
converging to $X^{*}$ from the right. While we have studied this
evolution only with the simplest term $(\Box\phi)^{2}$ in the action,
it is rather unclear whether one might improve the dynamical convergence
of the background field adding higher order terms $(\Box\phi)^{4}$,
$(\Box\phi)^{6}$, etc. along with higher than second derivative terms. 
Without higher than second derivative terms, the system would, in general, exhibit
oscillatory behaviour, provided $H$ is smaller than the amplitude of the oscillations. 
This can be understood by looking at the linearized 
action near $X^*$ with functions $Q(X)$ and $S(X)$ taken to be arbitrary and finite in the vicinity of 
the condensation point. Thus, the inclusion of the third, fourth, etc. derivative terms  into the 
Lagrangian seems to be necessary but complicates much the study of the 
dynamical evolution of the system.

We have studied two particular cases: the de Sitter background and
a matter dominated Universe. We have analyzed the evolution of the
homogeneous part of the field $\phi$ for the various
initial date. In the de Sitter case the overshoot does
not occur if $H_{inf}$ is greater then the scale $M$ (assuming, $\alpha\sim 1$). This is, however,
an unphysical regime of the EFT at the condensation
point since the powers of $\Box\phi$ higher than quadratic contribute
more and more to the action. This contribution increases consequently with the
power of the operator. Still, it is interesting to analyze the behaviour
of the truncated theory. It turns out, that
simple truncation Eq.~(\ref{eq:action one}) works well if $H\gtrsim M$.
If one wants to reconcile with the EFT at the condensation
point one need to assume the opposite. In that case, when $H\lesssim M$,
this truncation is not possible, which manifests itself in the oscillations
of the background homogeneous solution around the condensation point. 

In the matter dominated case the system always overshoots the attractor
point for any choice of the initial conditions. But there is a finite
time before the overshoot occurs. This time depends on how late the
effective Lagrangian starts to describe the dynamics of the system.
If the effective Lagrangian description of the low energy physics
becomes valid at the late times of the Universe evolution, the system
may stay in healthy hyperbolic regime for a period of time, which
may well exceed the present age of the Universe. In the end we summarize
our results and point out some future directions which are yet to
be explored.

\section{Formation Of The Ghost Condensate}

For a start we will consider the dynamics of the scalar field with
a purely kinetic Lagrangian and action given by \begin{equation}
S_{\phi}=M^{4}\int d^{4}x\sqrt{-g}P(X),\label{eq:Action Zero}\end{equation}
 where $M$ is some characteristic scale.
The equation of motion is \begin{equation}
\frac{1}{\sqrt{-g}}\partial_{\mu}\left[\sqrt{-g}P'(X)\partial^{\mu}\phi\right]=0,\label{eq:Equation of motion}\end{equation}
 It is easy to verify that following discussion is not sensitive to
the choice of $P(X)$ if this function has following properties: 

\begin{itemize}
\item $P(X)=P(0)-X+O(X^{2})$ near $X=0$
\item $P(X)$ has minimum at some $X^{*}\neq0$
\item The energy density at $X=X^{*}$ is zero 
\end{itemize}
For specific calculations we will be choosing $X^{*}=1$ and function
$P(X)=\frac{1}{2}(X-1)^{2}$.
In the spatially flat Friedmann universe with the interval\begin{equation}
ds^{2}=dt^{2}-a^{2}(t)\, d\mathbf{x}^{2},\end{equation}
 the Eq.~(\ref{eq:Equation of motion}) reads as \begin{equation}
\partial_{t}\left[P'(X)\partial_{t}\phi\right]+3HP'(X)\partial_{t}\phi-\frac{1}{a^{2}}\partial_{i}\left[P'(X)\partial_{i}\phi\right]=0,\label{eq:Equation of motion Friedmann}\end{equation}
 where $H=\dot{a}/a$ is a Hubble parameter, $a(t)$ is the scale
factor. The spatial part of Eq.~(\ref{eq:Equation of motion Friedmann})
is suppressed by growing $a(t)$ and vanishes as universe expands.
Thus, the solutions of Eq.~(\ref{eq:Equation of motion Friedmann})
eventually become nearly homogeneous in the cosmological reference
frame.

There are two attractors since\begin{equation}
P'(X)\dot{\phi}=\frac{const}{a^{3}(t)}\rightarrow 0~~ 
 {\rm as}~~t\rightarrow+\infty,\end{equation}
 namely $X=0$ and $X^{*}$ where $P'(X^{*})=0$. The solution converges
to either of them depending on the initial conditions (see arrows
on Fig.~1). The first attractor is unstable due to production of
the regular particles out of vacuum through the gravitational coupling
to ghosts \cite{Cline}. We will assume only those initial conditions
for $X$ which satisfy $X>X^{*}$. 

The sound speed for cosmological fluid with the action (\ref{eq:Action Zero})
is given by \cite{M}\begin{equation}
C_{s}^{2}\equiv\frac{P'(X)}{\varepsilon'(X)},\end{equation}
 and vanishes at $X=X^{*}$. Here $\varepsilon(X)=2XP'(X)-P(X)$ is
the corresponding energy density for the action (\ref{eq:Action Zero}).
We are free to choose the energy density to be equal to zero at $X^{*}$. It is easy
to check that $C_{s}^{2}>0$ $(C_{s}^{2}<0)$ to the right (left)
of $X^{*}$. Thus the theory is classicaly and quantum mechanically
stable only for $X>X^{*}$ (shaded region on Fig.~1).\begin{figure}[thb] \begin{center} \label{Pressure} \psfrag{X}[tr]{\small$X$} 

\psfrag{P}[tr]{\small$P(X)$} 

\psfrag{C}[tr]{\small$C^{2}_{s}$} 

\psfrag{0}[r]{\small$0$} 

\psfrag{X1}[tl]{\small$X^*$} 

\psfrag{S}[]{stable}

\psfrag{N}[r]{classicaly unstable} 

\psfrag{M}[r]{unstable}

\psfrag{QFT}[l]{QFT}

\psfrag{X2}[tr]{\small$X_{d}$}

\includegraphics[width=3in,angle=0]{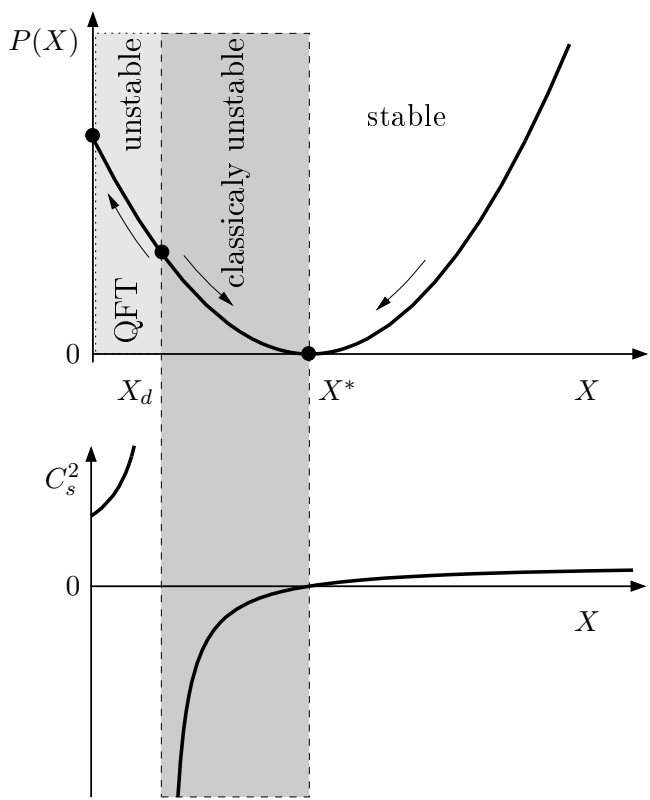} 

\caption{On the first graph the shaded regions are regions where the sign in front of $\dot\pi^2$ is negative. There is a point $X_d$ between $X=0$ and $X=X^*$ where the sound speed $C^2_s$ is divergent. Unshaded region is the one where the sign in front of $\dot\pi^2$ is positive along with $C^2_s$.} 

\end{center} 

\end{figure}

One can also expand the Lagrangian of (\ref{eq:Action Zero}) at some
arbitrary background value of $X$ (which we take sufficiently close
to $X^{*}$) in Minkowsky space up to quadratic order in $\pi$ which
is defined through $\phi=M^{2}\sqrt{2X}t+\pi$:\begin{equation}
\mathcal{L}=M^{4}\left\{ \left[2XP''(X)+P'(X)\right]\dot{\pi}^{2}-P'(X)(\nabla\pi)^{2}\right\} \label{eq:Lagrangian PI}\end{equation}

It is easy to see that the field theory described by Lagrangian (\ref{eq:Lagrangian PI})
is again classicaly stable if $X>X^{*}$, since the signs of coefficients
in front of $\dot{\pi}^{2}$ term and $(\nabla\pi)^{2}$ are opposite.
At $X_{d}<X<X^{*}$ these signs are the same and the theory is classicaly
unstable because the equation of motion is of the elliptic type. 
It is worthwile noting that two regions of classical stability $X<X_d$ and $X>X^*$ on the $X$ axis 
(where the system is hyperbolic) are disconnected. 
Thus the casual evolution from one stable regime to another is classicaly forbidden. 
Therefore, for the theory without higher derivatives terms in the action, 
the field described by the "ghost" Lagrangian cannot classicaly 
evolve to the right neighborhood of $X^*$. 
\\

Having described how cosmological kinetic condensate appears, we next
modify the action (\ref{eq:Action Zero}) adding the simplest general
covariant higher derivative term\begin{equation}
S_{\phi}=\int d^{4}x\sqrt{-g}\left[M^{4}P(X)-\frac{\alpha^{2}(\Box_{g}\phi)^{2}}{2M^{2}}\right]\label{eq:action one}\end{equation}
 which induces nontrivial modification of the graviton propagator. 

We will now turn to the question of how does $\alpha$-term effects
the picture of the condensate formation reviewed above. In the next
section we derive the equation of motion in the homogeneous case for
any function $H(t)$. Then we solve exactly the linearized equation
of motion for two cases :

\begin{itemize}
\item $H=const$, de Sitter Universe
\item $H=\frac{2}{3t}$, i.e. matter dominated Universe
\end{itemize}
In the case of a large Hubble parameter the overshoot never occurs for any
reasonable choice of the initial conditions $(\dot\phi_0,\ddot\phi_0,
\dddot\phi_0)$. 
If $H\ll M$ the homogeneous solution is oscillatory.
 In both cases we will assume that the energy density due to the condensate
is always a tiny fraction of the total energy density, so that it
has no effect on the value of the Hubble parameter.  
This is not restrictive at all since $M/M_{pl}\ll 1$. At the same time we  
do not have to worry about condition $\rho_{tot}>0$. This allows us to 
consider any set of initial values for all derivatives even such that 
$\varepsilon_{0}<0$ .

\section{Does the system overshoot $X^{*}$ point?}

In order to avoid any possible confusion in the rest of this paper
all the units are scaled with appropriate power of $M$ and are all
dimensionless unless it is specified otherwise. 
The equation of motion in the homogeneous case is: 
\begin{equation}\begin{array}{c}
a^{-3}\partial_{t}\left(a^{3}P'(X)\dot{\phi}\right)=\\~~\\-\alpha^{2}\left[\phi^{(4)}+
6H\dddot{\phi}+\ddot{\phi}\left(6\dot{H}+9H^{2}\right)+\dot{\phi}\left(9H\dot{H}+3\ddot{H}\right)\right] 
\label{eq:Box Eq.of Motion(FRW)}
\end{array}
\end{equation}
The first part is easily recognizable from the equation of motion
derived for $P(X)$ Lagrangian. The second comes from the $\alpha$-term.

In order to investigate the possibility of overshoot we will go to
some point where $X>X^{*}$ choosing some initial values for the second
and third derivatives of $\phi$. For $P(X)$ we will use the form:
$P(X)=\frac{1}{2}(X-1)^{2}$.

\subsection{Ghost Condensation In The de Sitter Phase}

In this section we assume that there is a phase transition which occurs
before or during inflation leading to the the effective field theory
of \cite{ah}. Introducing a new variable $y\equiv\dot{\phi}$ we
can simplify the equation of motion to

\begin{equation}
\dddot{y}+6H\ddot{y}+9H^{2}\dot{y}+\alpha^{-2}\left(\varepsilon'\dot{y}+3yHP'\right)=0.\label{eq:Eq of Motion in y}\end{equation}

Note that $X^{*}=1$ corresponds to $\dot{\phi}^{*}=\sqrt{2}$. Inserting
the expression for $P(X)=\frac{1}{2}(X-1)^{2}$ and linearizing the
equation around attractor point as $y=\sqrt{2}+\eta$, where $\eta$
is a small deviation from the attractor point, we obtain the following
equation: \begin{equation}
\dddot{\eta}+6H\ddot{\eta}+\left(9H^{2}+\frac{2}{\alpha^{2}}\right)\dot{\eta}+\frac{6H\eta}{\alpha^{2}}=0.\end{equation}
This equation is easy to analyze. The characteristic equation reads
as

\begin{equation}
\lambda^{3}+6H\lambda^{2}+\left(9H^{2}+\frac{2}{\alpha^{2}}\right)\lambda+\frac{6H}{\alpha^{2}}=0.\end{equation}
The roots can be easily found: 
\begin{equation}
\lambda_{1}=-3H,~~\lambda_{\pm}=-\frac{3}{2}H\left[1\pm\sqrt{1-\frac{8}{\left(3H\right)^{2}\alpha^{2}}}\,\right],\label{eq:Roots}\end{equation}
Note that all roots are negative. Parameter $\alpha$ is generally
assumed to be of order of unity. Since $H$ is scaled in units of $M$
then, for example, in the case of standard chaotic inflation and for
$M\sim1$ GeV the first two roots are of order of $|\lambda_{1,+}|\sim10^{13}$.
The third root can be Taylor expanded: $\lambda_{-}\simeq-\frac{2}{3} \alpha^{-2} H^{-1}$
and is very tiny in this case. 

It is easy to check that $\ddot{\phi}$ and $\dddot{\phi}$ decay
as $\exp(-3HT)$, while $\dot{\phi}$ picks up the smallest root:
$\dot{\phi}-\sqrt{2}\sim\exp(\lambda_{-}T)$. Here $T$ is the dimensionless
time: $T=Mt$. In this case the value of the first derivative gets
stuck at its initial value for a very long period. It is easy to estimate
that inflation should last as long as $N^{*}$ e-folds in order to
get $\dot{\phi}$ very close to the attractor point, where $N^{*}=(H/M)^{2}$.
Although, we do not know for how long a real inflation lasted the
number of e-foldings $N^{*}$ for the case when $H\sim10^{13}$ GeV
and $M\sim1$ GeV is of order of $10^{26}$ does not look very natural.
However if $H/M$ is not too big $\dot{\phi}$ may come sufficiently
close to the value $\dot{\phi}=\sqrt{2}$. This is illustrated on the Figs.~2, 3. 
\begin{figure}[thb] \begin{center} \label{de Sitter light} 

\psfrag{X}[bl]{\small$X$} 

\psfrag{T}[bl]{\small$T$}

\psfrag{H1}[bl]{\small$H/M=10^{13}$}

\psfrag{H2}[bl]{\small$H/M=10^{2}$}

\includegraphics[width=3in,angle=0]{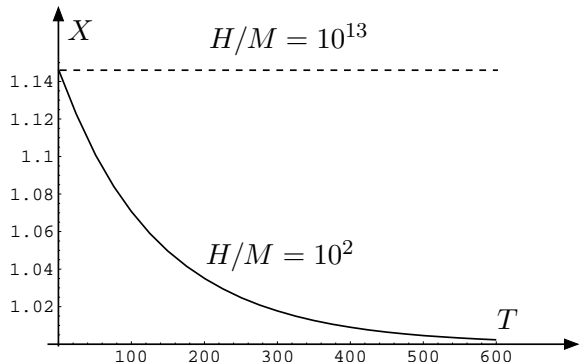} 

\caption{Here we plot evolution of X in de Sitter case for initial values  $\dot\phi=\sqrt{2}+0.1,\ddot\phi=0.1,\dddot\phi=0.1$. One can see that if ${H\over M}=10^2$ the value of $X$ is efficiently driven towards $X^*=1$. However for ${H\over M}=10^{13}$ on the same time scale the value of $X$ looks frozen.} 

\end{center} 

\end{figure}

\begin{figure}[thb] \begin{center} \label{de Sitter light Deriv} 

\psfrag{X}[tl]{\small$\ddot\phi$} 

\psfrag{X1}[tl]{\small$\dddot\phi$}

\psfrag{T}[bl]{\small$T$} 

\includegraphics[width=3in,angle=0]{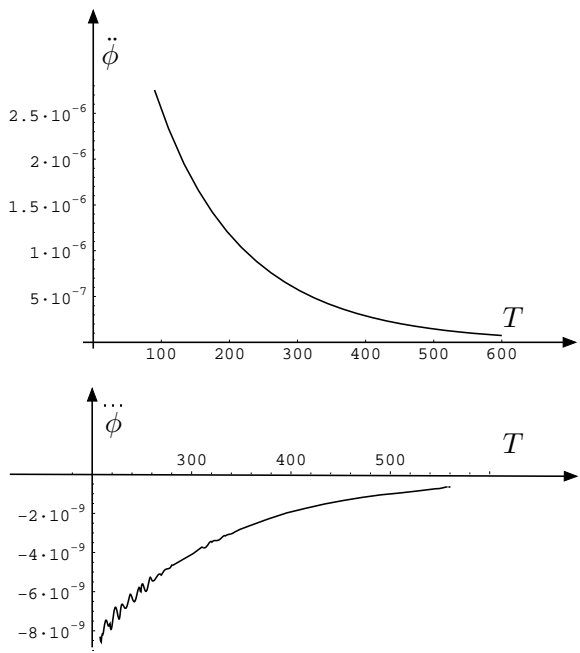} 

\caption{Here one can see that higher derivatives of $\phi$ in de Sitter case are quickly converging to zero from their initial values $\ddot\phi=0.1,\dddot\phi=0.1$. ${H\over M}$ was chosen to be $10^2$ and $\dot\phi=\sqrt{2}+0.1$.} 

\end{center} 

\end{figure}

We can now conclude that ghost condensation occurs efficiently during
de Sitter stage only if $H$ is not much larger then scale $M$. Thus
we are forced to assume that either scale $M$ is very high or we
need low scale inflation, for which there are several examples can
be found \cite{low} which allows to consider smaller values of $M$.
Otherwise there could be no condensation during inflation. On the
other hand we can consider outcome of high scale inflation which is:
$X_{0}>X^{*}$ and $\ddot{\phi_{0}}=\dddot{\phi_{0}}=0$. Continuing
into following short matter dominated stage during preheating and
then into radiation dominated stage we can investigate what happens
if we start with these particular initial conditions. 

In the de Sitter universe with a Hubble parameter  $H\ll M$ the solution
oscillates around condensation point. This is shown on Fig.~4 for
two values of $H$. Such behaviour can also be deduced from Eq.
(\ref{eq:Roots}). \begin{figure}[thb] \begin{center} 

\psfrag{X}[bl]{\small$X$} 

\psfrag{t}[bl]{\small$T$}

\psfrag{H1}[bl]{\small$H/M=10^{-2}$}

\psfrag{H2}[bl]{\small$H/M=10^{-1}$}

\includegraphics[width=3in,angle=0]{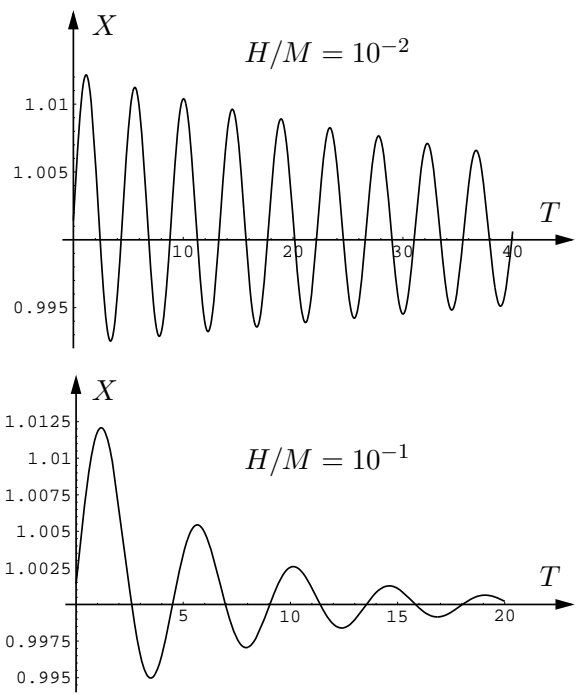} 

\caption{Here we plot evolution of X in de Sitter case for initial values  $\dot\phi=\sqrt{2}+0.001,\ddot\phi=0.01,\dddot\phi=0.001$. Hubble parameter is chosen to satisfy $H\ll M$. Classical evolution of a background field is oscillatory around condensation point.} 

\end{center} 

\end{figure}In the next subsection we consider evolution of $X$ in the matter
dominated case for different initial values of $X_{0}>X^{*}$ and
higher derivatives of $\phi$. It is worthwile noting that the initial values considered in this case are in 
the region of validity of the EFT. Nevertheless the system evolves to a regime which has the oscillations with
frequency $\sim M$ that is outside of the regime of validity of EFT. Due to the Hubble dumping of these oscillations one can 
hope that the instability is not so dangerous. However, this shows the limitations in the use of the truncated action
for a viable description of the ghost condensation.

\subsection{Ghost Condensation In the Matter Dominated Universe}

For $H=\frac{2}{3t}$ it worth noting that some miraculous cancellations
occurs in the equation of motion, namely coefficients in front of
$y$ and $\dot{y}$ turn out to be zero: \begin{equation}
\dddot{y}+6H\ddot{y}++\frac{1}{\alpha^{2}}\left(\varepsilon'\dot{y}+3yHP'\right)=0.\label{eq:Eq of Motion for Matter}\end{equation}
Linearizing this equation in the neighborhood of the {}``attractor''
$X^{*}=1$ with $P(X)=\frac{1}{2}(X-1)^{2}$ and $H(t)=\frac{2}{3}t^{-1}$
we obtain\begin{equation}
\dddot{\eta}+\frac{4}{t}\left(\ddot{\eta}+\frac{\eta}{\alpha^{2}}\right)+\frac{2}{\alpha^{2}}\dot{\eta}=0.\label{eq:Linearized Eq of Motion for MatterDom}\end{equation}
By rescaling $\tau\equiv t\sqrt{2}/\alpha$ this equation can be simplified
even further:\begin{equation}
\dddot{\eta}+\frac{2}{\tau}\left(2\ddot{\eta}+\eta\right)+\dot{\eta}=0\end{equation}
The general solution of this equation is
\begin{equation}
\eta(\tau)=\frac{1}{\tau^{2}}\left[c+\left(a\cos\tau+b\sin\tau\right)\right]+{1\over\tau}\left(a\sin\tau-b\cos\tau\right).
\label{eq:Solution of Matter Dominant}
\end{equation}

Now we will provide the proof that sooner or later independently of
the initial conditions $\eta(\tau)$ will take negative value signalling
overshoot. 

First case we will consider is $\dot{\phi}=\sqrt{2}+\eta_{0}$, $\ddot{\phi}=0$,
$\dddot{\phi}=0$. The solution for $\eta(\tau)$ is 
\begin{equation}
\eta(\tau)=\eta_{0}[\frac{\tau_{0}^{2}-1}{\tau^{2}}+\frac{1}{\tau^{2}}\cos(\tau-\tau_{0})+\frac{1}{\tau}\sin(\tau-\tau_{0})].
\end{equation}
 Note again, that we are working with dimensionless variables, all
in units of $M$. Inevitability of overshoot is obvious in this case
because first two terms scale as $\frac{1}{\tau^{2}}$ and eventually
become negligible with respect to the third one. But $\sin(\tau-\tau_{0})$
is oscillating function which takes both positive and negative values.
This proves that starting anywhere near the attractor point with $\ddot{\phi}=0$
and $\dddot{\phi}=0$ there is overshoot. Note that this set of initial
conditions corresponds to the outcome values in the end of inflation.
The coefficient in front of a $(\nabla\pi)^{2}$ in the Lagrangian
becomes negative and the system moves into instability region. It
is easy to estimate time within overshoot happens. Assuming $\tau_{0}\gg1$
the estimate follows from 

\begin{equation}
\frac{\tau_{0}^{2}}{\tau^{2}}\sim\frac{1}{\tau},\end{equation}
and overshoot time is of order 

\begin{equation}
t_{over}\sim\frac{M}{H_{*}^{2}},
\label{eq:t_over estimate}
\end{equation}
where $H_{*}$ is the value of the Hubble constant at a time the system
starts to evolve. Here we have returned to the dimensionfull parameters
$M$ and $H_{*}$. Making simple estimates for $M=1$ GeV and $H^{*}=10^{3}$
GeV we see that overshoot occurs within a time $t_{over}\sim10^{-9}$
GeV$^{-1}$. If we assume that at this time the Universe is matter
dominated, we see that first overshoot occurs within a time much smaller
then one Hubble time. Thus we can conclude that early matter dominated
epoch destabilizes $\pi$ excitations very quickly.

On the other hand we can make an estimate assuming that phase transition
occurred at, for example, $t\sim10^{6}$ years, i.e. at the beginning
of the second matter domination epoch, which follows radiation dominated
era. The value of $H^{*}\sim10^{-13}sec^{-1}$, $M=1$ MeV corresponds
to $10^{-22}sec$. The overshoot time is $t_{over}\sim10^{41}$years.
This time is enormously large compared to the present age of the Universe. 

Therefore, we can conclude that only if the phase transition occurs
late enough in the history of the Universe the excitations of $\pi$
stay healthy for a very long time. However, early phase transition
leads to quick instability, and thus, an implausible variant.

\begin{figure}[thb] 

\begin{center} 

\label{Matter} 

\psfrag{X}[tl]{\small$X$} 

\psfrag{T}[bl]{\small$T$} 

\includegraphics[width=3in,angle=0]{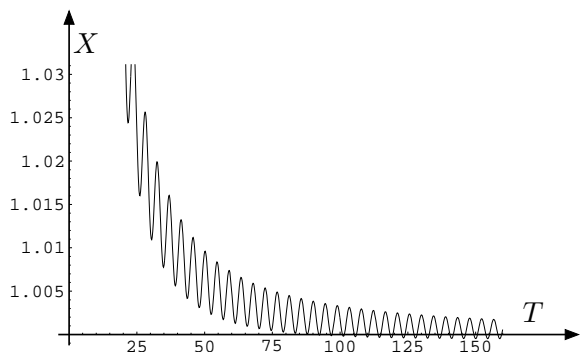} 

\caption{Here we plot evolution of X in a matter dominated case for initial values: $\tau_0=10$, $\dot\phi=\sqrt{2}+0.1,\ddot\phi=0,\dddot\phi=0$. 
These initial values can be considered as an outcome of high scale inflation.
First overshoot occurs at a time of order of $\tau_0^2$, as expected from Eq. (\ref{eq:t_over estimate}).}

\end{center} 

\end{figure}\begin{figure}[thb] \begin{center} 

\psfrag{X}[tl]{\small$X$} 

\psfrag{T}[bl]{\small$T$} 

\includegraphics[width=3in,angle=0]{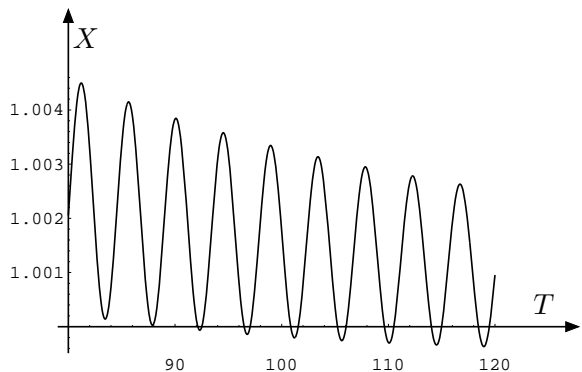} 

\caption{On this graph we simply show a part of the Fig.5 in more details. 
One can see that periodically the value of $X$ start to cross the $X=1$ point towards lower values. 
This means that the sign of the $(\nabla\pi)^2$ term changes periodically.} 

\end{center} 

\end{figure}

\begin{figure}[thb] \begin{center}

\psfrag{X}[tl]{\small$X$} 

\psfrag{T}[bl]{\small$T$} 

\includegraphics[width=3in,angle=0]{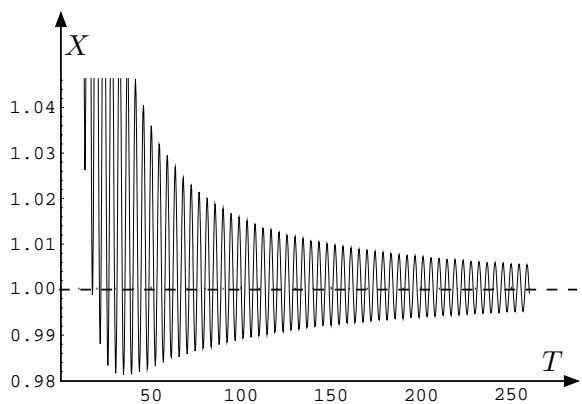} 

\caption{This graph illustrates overshoot for the matter dominated case. The initial values are taken to be $\tau_0=10$,
$\dot\phi=\sqrt{2}+0.1,\ddot\phi=0.1,\dddot\phi=0.1$. 
This picture is generic. 
The time of the first overshoot can be estimated from Eq. (\ref{eq:overtime}) if one computes coefficients $a$ and $b$.} 

\end{center} 

\end{figure}

In general, one can see that for any choice of $\ddot{\phi}_{0}$
and $\dddot{\phi}_{0}$ one can rewrite the general solution in the
form

\begin{equation}
\eta(\tau)=\frac{c}{\tau^{2}}+\frac{\sqrt{a^{2}+b^{2}}}{\tau^{2}}\cos(\tau-\delta)+\frac{\sqrt{a^{2}+b^{2}}}{\tau}\sin(\tau-\delta),\end{equation}
where $\cos(\delta)\equiv a/\sqrt{a^{2}+b^{2}}$. Because $\sin(\tau-\delta)$
suppressed by one power of time, while the rest of the terms are suppressed
by $\tau^{2}$, it eventually starts to dominate. Thus, overshoot
is inevitable for an arbitrary choice of $\ddot{\phi}(\tau_{0})$
and $\dddot{\phi}(\tau_{0})$. The upper bound on the time duration
within which the system can stay in stable regime is defined by the
condition that the first two terms are larger then the second. This
leads roughly to the condition 

\begin{equation}
\tau_{max}\sim\frac{\eta_{0}\tau_{0}^{2}}{\sqrt{a^{2}+b^{2}}},\label{eq:overtime}\end{equation}
where coefficients $a$ and $b$ depend on initial values of derivatives
of $\phi$. There is special choice of these when $a=b=0$. Then the
system does not overshoot $X^{*}$ point. However, this choice has
zero measure. Moreover, this is true only in linearized approximation.
In general, estimate (\ref{eq:overtime}) gives correct upper bound
on the time during which condensate is in stable region.

\section{Summary}

We have shown that the dynamical mechanism of spontaneous Lorentz violation 
(dubbed as "ghost condensation'') 
 is somewhat troubled in the simplest case  of the general covariant
Lagrangian which has second derivative operator. Oscillatory behaviour 
of the homogeneous part of the field $\phi$ leads to an overshoot of the 
system into unstable elliptic regime in most cases. This does not occur  
during the de Sitter phase if the Hubble parameter is larger then the UV 
cutoff of the effective field theory. At very small values of $H/M$ the 
solution of the classical equations of motion is oscillatory again. 
In the matter domination case the theory may be stable for a long 
period depending on the scale $M$ and the value of the Hubble 
parameter $H^{*}$ at the time, when the effective field theory 
description of the evolution of the "ghost" field $\phi$ becomes valid. 
However, eventually the system moves into a classicaly unstable region. 
Therefore, the simplest truncated action may have a limited use.

We would like to make the following cautionary remark. 
The overshoot in our case has oscillatory nature. 
An oscillatory instability is milder then the usual one. 
It is very interesting to understand up to what degree 
this overshoot is implausible. For example, in the de Sitter phase with 
a small Hubble parameter $H\ll M$, having the initial values fine tuned extremely close to 
$X^*$, the instability may not have enough time to develop. Because the 
characteristic timescale of this instability may be smaller than the period of one 
oscillation \cite{MukoP}.
Nevertheless these initial values, in general, do not yield "smooth''  mechanism of the 
ghost condensation.    

There are several other questions left. First, the phenomenological
bounds on the scale $M$. As we mentioned before there are two scenarios.
If the first one is realized one have to look for the signatures of
that in CMBR spectrum. The impact of ghost condensation during inflation
depends on the contribution of the ghost condensate to the total energy
density. In the case which we have looked at, this effect is probably
negligible. 

The other interesting question is if quantum fluctuations can take the system 
from near $X>X^*$ into the state with $X<X^*$. If this happens in some 
macroscopic bubble (with the initial volume $V\gg M^{-3}$) the field inside the bubble has a 
peculiar classical evolution studied in details in
\cite{rz}. It is interesting, that this analysis may
well be applied to a {}``domain wall'' like objects, which may form because of $Z_{2}$ symmetry:
$\dot{\phi}\rightarrow-\dot{\phi}$.
However, probability of such quantum fluctuations in both cases 
is hard to estimate and is 
likely to be suppressed.

If the evolution of the underlying theory is described by the action \ref{eq:action one}
starting from a small value Hubble parameter, the overshoot occurs rather late.
In that case it is interesting to discuss implications of the existence of the condensate
and $\pi$'s theory at present. Some bounds on $M$ were already discussed
in the literature \cite{ah,frolov,Muko}. Interesting phenomenology was
also discussed in \cite{dub, Muko}.

It would be interesting to find out if any new gravitational effects
could be observed from a reference frame which moves fast with respect
to the cosmological reference frame. By fast we mean faster than the
velocity associated with the Jeans instability discussed in \cite{ah}.
This is interesting simply because every object in the Universe is
moving with respect to the cosmological reference frame. This velocity
is generically of order of $10^{-3}$ in units of the speed of light.
Thus, any potential observer would find itself on a fast moving source.
We will try to improve on this and some other questions in \cite{future}. 

In the end we note that the problem of overshoot discussed in this
paper is solely due to higher order derivative operator $(\Box\phi)^{2}/M^{2}$.
There is no overshoot if one considers $P(X)$ only. But then, there
is no modification of gravity in the infrared either. The case reduces
to some specific k-essence model. 

%%%%%%%%%%%%%%%%%%%%%%%%%%%%%%%%%%%%%%%%%%%%%%%%%%%%%%%%%%%%%%%%%%%%%%
\section*{Acknowledgments}
We would like to thank Sergey Dubovsky, Slava Mukhanov and Sergei Winitzki
for illuminating discussions. We are also grateful to Shinji Mukohyama and Nima 
Arkani-Hamed for valuable comments for the revised version of this paper.
We are specially thankful to Valentine
Zakharov who had originally inspired the interest to this subject
and carried on a lot of discussion concerning ghost condensation scenario.

\vspace*{2cm}

%%%%%%%%%%%%%%%%%%%%%%%%%%%%%%%%%%%%%%%%%%%%%%%%%%%%%%%%%%%%%%%%%%%%%%
\section*{References} %%%%%%%%%%%%%%%%%%%%%%%%%%%%%%%%%%%%%%%%%%%%%%%%
%%%%%%%%%%%%%%%%%%%%%%%%%%%%%%%%%%%%%%%%%%%%%%%%%%%%%%%%%%%%%%%%%%%%%%


\begin{thebibliography}{99}
\bibitem{ah}N. Arkani-Hamed, Hsin-Chia Cheng, M. A. Luty, S. Mukohyama, ''Ghost
Condensation And A Consistent Infrared Modification Of Gravity,\char`\"{}
JHEP 0405:074 (2004) {[}Preprint hep-th/0312099{]}
\bibitem{k_inf}C. Armendariz-Picon, T. Damour, and V. Mukhanov, ''K-Inflation,\char`\"{}
Phys.Lett. B \textbf{458}, 209 (1999) {[}Preprint hep-th/9904075{]}; 
\bibitem{k_ess1}C. Armendariz-Picon, V. Mukhanov, and P. J. Steinhardt, ''Essentials
Of K-Essence,\char`\"{} Phys.Rev. \textbf{D} 63, 103510 (2001) {[}Preprint astro-ph/0006373{]}
\bibitem{k_ess2}C. Armendariz-Picon, V. Mukhanov, P. Steinhardt, ''A Dynamical Solution
To The Problem Of A Small Cosmological Constant And Late Time Cosmic
Acceleration, Phys.Rev.Lett.\textbf{85}, 4438 (2000) {[}Preprint astro-ph/0004134{]} 
\bibitem{k_ess3}T. Chiba, T. Okabe, M. Yamaguchi, ''Kinetically Driven Quintessence'',
Phys.Rev.D62:023511 (2000) {[}Preprint  astro-ph/9912463{]} 
\bibitem{br_w}G. R. Dvali, G. Gabadadze, and M. Porrati, ''4D gravity on a brane
in 5D Minkowski space,\char`\"{} Phys. Lett. B \textbf{485}, 208
(2000) {[}Preprint hep-th/0005016{]}; 
\bibitem{br_w1}G. R. Dvali and G. Gabadadze, ''Gravity on a brane in infinite-volume
extra space,\char`\"{} Phys. Rev. D \textbf{63}, 065007 (2001) {[}Preprint hep-th/008054{]}; 
\bibitem{br_w2}C. Deffayet, G. R. Dvali, and G. Gabadadze, ''Accelerated universe
from gravity leaking to extra dimensions,\char`\"{} Phys. Rev. D
\textbf{65}, 044023 (2002) {[}Preprint astro-ph/0105068{]}
\bibitem{mod}I. T. Drummond, ''Bimetric Light-Cone Theory Of Gravity,\char`\"{}
{[}Preprint gr-gc/9908058{]}; 
\bibitem{mod1}T. Jacobson and D. Mattingly, ''Gravity with a dynamical preferred
frame,\char`\"{} Phys. Rev. D \textbf{64}, 024028 (2001) {[}Preprint gr-gc/0112012{]}; 
\bibitem{mod2}K. Freeze and M. Lewis, ''Cardassian Expansion: a Model in which
the Universe is Flat, Matter Dominated, and Accelerating,\char`\"{}
Phys. Lett. B \textbf{540}, 1 (2002){[}Preprint astro-ph/0201229{]}
\bibitem{ah2}N. Arkani-Hamed, Paolo Creminelli, Shinji Mukohyama, Matias Zaldarriaga,
''Ghost Inflation\char`\"{}, JCAP 0404:001 (2004) {[}Preprint hep-th/0312100{]} 
\bibitem{other}V. A. Kostelecky, R. Lehnert, M. J. Perry, {}``Spacetime - Varying
Couplings And Lorentz Violation'', Phys.Rev.D68:123511 (2003), {[}Preprint 
astro-ph/0212003{]}; O. Bertolami, R. Lehnert, R. Potting, A. Ribeiro
{}``Cosmological Acceleration, Varying Couplings, And Lorentz Breaking''.
Phys.Rev.D69:083513 (2004) {[}Preprint astro-ph/0310344{]}
\bibitem{vlad}V. S. Vladimirov, {}``Equation of Mathematical Physics'', Nauka,
Moscow, 1983, pp. 81-83.
\bibitem{M}J. Garriga, V. F. Mukhanov,~{}``Perturbations in k-inflation'' ~Phys.Lett.
B \textbf{458}, 219 (1999), {[}Preprint hep-th/9904176{]}.
\bibitem{Cline}J. M. Cline, S. Y. Jeon and G. D. Moore,  {}``The phantom menaced:
Constrains on low-energy effective ghosts'', {[}Preprint hep-th/0311312{]}; 
S. M. Carroll, M. Hoffman and M. Trodden, 
"Can the dark energy equation-of-state parameter  w  be less than -1?", 
Phys. Rev. D 68 023509 (2003), [Preprint astro-ph/0301273].
\bibitem{low}A. Linde, ''Hybrid Inflation\char`\"{}, Phys.Rev.D49:748-754 (1994)
{[}Preprint astro-ph/9307002{]}; D. Lyth, D. Wands, ''Generating The
Curvature Perturbation Without An Inflation\char`\"{}, Phys.Lett.B524:5-14
(2002) {[}Preprint hep-ph/0110002{]} 
\bibitem{rz}D. Krotov, C. Rebbi, V. Rubakov, V. Zakharov, {}``Holes In The Ghost
Condensate'', {[}Preprint hep-ph/0407081{]} 
\bibitem{frolov}A. Frolov, ''Accretion Of The Ghost Condensate By Black Holes\char`\"{},
{[}Preprint hep-th/0404216{]} 
\bibitem{MukoP}S. Mukohyama, private communication
\bibitem{Muko}S. Mukohyama, `` Black Holes In The Ghost Condensate'', Preprint hep-th/0502189;
\bibitem{dub}S. Dubovsky, {}``Star Tracks In The Ghost Condensate'', JCAP 0407:009
(2004), {[}Preprint: hep-ph/0403308{]}; M. Peloso, L. Sorbo, {}``Moving
Sources In A Ghost Condensate'', Phys.Lett.B593:25-32 (2004), [Preprint hep-th/0404005];
A. Krause, Siew-Phang Ng, ``Ghost Cosmology: Exact Solutions, Transition Between
Standard Cosmologies And Ghost Dark Energy/Matter Evolution'', [Preprint hep-th/0409241{]}
\bibitem{future}A. Anisimov, A.Vikman, \textit{in preparation}\end{thebibliography}
\end{document}